\documentclass[12pt,a4paper]{article}

\RequirePackage{graphicx}
\RequirePackage{mathptmx}      
\RequirePackage{flushend}
\RequirePackage[numbers,sort&compress]{natbib}

\usepackage{breakurl}
\usepackage{graphicx}
\usepackage{amsfonts}
\usepackage{amssymb}
\usepackage{amsbsy}
\usepackage{amsmath}
\usepackage{mathrsfs}
\usepackage{latexsym}
\usepackage{bm}
\usepackage{subfigure}
\usepackage{color}
\def\xx{\mathbf{x}}

\def\x{\mathbf{x}}

\def\rs{r_s}

\begin{document}

\title{Remarks on distinguishability of Schwarzschild spacetime and thermal Minkowski spacetime 
using Resonance Casimir-Polder interaction}

\author{ Chiranjeeb 
Singha\\
\emph{Department of Physical Sciences, Indian Institute of Science Education }\\
\emph{and Research Kolkata, Mohanpur - 741 246, WB, India}\\
\small{ cs12ip026@iiserkol.ac.in}}

\maketitle

\date{}
\begin{abstract}
 
One perceives same response of a single-atom detector when placed at a point outside the 
horizon in Schwarzschild spacetime to that of a static 
single-atom detector in thermal Minkowski spacetime. So one cannot distinguish 
Schwarzschild spacetime from thermal Minkowski spacetime by using a single-atom 
detector. We show that, for Schwarzschild spacetime, beyond a characteristic length scale which is 
proportional to the inverse of the surface gravity $\kappa$, the \emph{Resonance 
Casimir-Polder interaction (RCPI)} between two entangled atoms is characterized 
by a $1/L^2$  power-law provided the atoms are located close to the horizon. 
However, the \emph{RCPI} between two entangled 
atoms follows a $1/L$ power-law decay for the thermal 
Minkowski spacetime. Seemingly, it appears that the spacetimes can be distinguished from each other using the \emph{RCPI} behavior.
But our further exploration leads to the conclusion that the length scale limit beyond a characteristic value is not compatible with the local flatness of the spacetime.

\end{abstract}

\section{Introduction}

\emph{Casimir Polder interaction (CPI)} is a very interesting phenomenon that 
arises due to the vacuum fluctuations of quantum field. The effects of the 
\emph{CPI} are widely investigated in many branches of physics with fairly 
accurate experimental corroboration \cite{book:786807}. Successful efforts  have 
been made  to probe more complicated contexts like entanglement \cite{refId0}, 
spacetime curvature 
\cite{Tian:2014hwa,Tian:2016uwp,PhysRevA.88.064501}, Unruh effect 
\cite{PhysRevA.76.062114,PhysRevLett.113.020403,Rizzuto:2016ijj}, Hawking radiation of a black hole
\cite{Yu:2008zza} and to check 
thermal and nonthermal scaling in a black hole spacetime \cite{Menezes:2017akp}. 
It is shown that the background spacetime and the relativistic motion of 
interacting systems can modify the \emph{CPI} 
\cite{PhysRevA.76.062114,PhysRevLett.113.020403,Rizzuto:2016ijj,Tian:2014hwa, 
Tian:2016uwp,PhysRevA.88.064501,Yu:2008zza,Menezes:2017akp,Zhou:2017axh,PhysRevD.97.105030,PhysRevD.79.044027,PhysRevD.88.104003,Salton_2015}. So 
one can in principle extract the information about the background spacetime and 
the relativistic motion of interacting systems from the \emph{CPI} 
\cite{Tian:2016uwp}.

It is well known that the response of a single-atom detector when placed at a 
point outside the horizon in Schwarzschild spacetime  is the same as the 
response of a static single-atom detector in thermal Minkowski 
spacetime\cite{Singleton:2011vh,Smerlak:2013sga,ISRAEL1976107,ISRAEL1981204,Louko:1998dj, Hinton:1982, Birrell1984quantum, 
Hodgkinson:2014iua,Acquaviva:2011vq,Acquaviva:2013oaa,Sriramkumar:1994pb, 
book:250016}. So one cannot distinguish these two spacetimes by using a 
single-atom detector.

In this paper, we follow the method as used in \cite {Tian:2016uwp} where the 
\emph{RCPI} occurs when one or more atoms are in their excited states and 
exchange of real photons is involved between them  \cite{book:842958, 
book:1285244}. Our system is modeled as two entangled atoms which are coupled 
with a massless scalar field. Here we consider the atoms in Schwarzschild 
spacetime and in thermal Minkowski spacetime. We show that, for Schwarzschild spacetime, beyond a characteristic length scale which is 
proportional to the inverse of the surface gravity $\kappa$, the \emph{Resonance 
Casimir-Polder interaction (RCPI)} between two entangled atoms is characterized 
by a $1/L^2$  power-law provided the atoms are located close to the horizon. 
However, the \emph{RCPI} between two entangled 
atoms follows a $1/L$ power-law decay for the thermal 
Minkowski spacetime. Seemingly, it appears that the spacetimes can be distinguished from each other using the \emph{RCPI} behavior.
But our further exploration leads to the conclusion that the length scale limit beyond a characteristic value is not valid for Schwarzschild spacetime. 

This article is organized as follows. In the section \ref{Time evolution of two 
atoms}, we follow the method as used in \cite{Tian:2016uwp}. Here we apply the 
open quantum system approach \cite{PhysRevA.70.012112,book:9622} to get the effective 
Hamiltonian of the two atoms which interact weekly with a massless scalar field. 
Using this  Hamiltonian, we can compute the shifts of the energy level of the 
symmetric state and the antisymmetric state of the two-atom system. It is shown 
that these shifts of the energy level are related to the two-point functions 
which can be calculated along the trajectories of the atoms and thus they depend 
on the spacetime background. As a result, these shifts of the energy level are 
different in different spacetimes.

In the section \ref{Schwarzschild spacetime}, we compute the two-point functions 
and the shifts of the energy level of the symmetric state and the antisymmetric 
state for two static atoms in Schwarzschild spacetime when the atoms are located 
close to the horizon. We show that beyond a characteristic length scale which is 
proportional to the inverse of the surface gravity $\kappa$, the \emph{RCPI} 
between two entangled atoms is characterized by a $1/L^2$ power-law decay for 
this spacetime. But the length scale limit beyond a characteristic value is not valid for Schwarzschild spacetime.

In the section \ref{Minkowski spacetime}, we compute the \emph{RCPI} between two 
entangled atoms for thermal Minkowski spacetime. We show that the \emph{RCPI} is 
always characterized by a $1/L$ power-law decay for this spacetime.

\section{Time evolution of two atoms}\label{Time evolution of two atoms}

Here we follow the method as used in \cite{Tian:2016uwp} and throughout the 
paper, we use \emph{natural units} where $c=\hbar=1$. For convenience, we adopt 
the same notations as in \cite{Tian:2016uwp}. Now we consider two atoms which 
are mutually independent and identical and these atoms interact weekly with a 
massless scalar field. Each atom has two internal energy levels correspond to 
two eigenstates, $+\frac{1}{2}\omega_0$ for the excited state, $|e\rangle$, and 
$-\frac{1}{2}\omega_0$ for the ground state, $|g\rangle$, respectively. The 
total Hamilton of the system is given by \cite{Tian:2016uwp},
\begin{equation}
H=H_{A}+H_{F}+H_{int}~,
\end{equation}
where $H_{A}=\frac{1}{2}\omega_0\sigma_3^{(1)}+\frac{1}{2}\omega_0\sigma_3^{ 
(2)} $ is the total Hamilton of two isolated atoms where superscript (1 or 2) 
labels the atom number, $\sigma_j$ with $j \in \{1,2,3\}$ are Pauli matrices.  Here $|e\rangle$ and $|g\rangle$ are eigenstates of $\sigma_3$ and
$H_{F}=\sum_k\omega_k
a^+_k a_k\frac{d t}{d \tau}$ is the free Hamiltonian of the massless field  where $a^+_k$ and $a_k$ are the creation and annihilation operators of the massless scalar field with linear dispersion relation $\omega_k=|k|$     \cite{PhysRevLett.113.020403} and $H_{int}$ is the 
field-atom interaction term which is assumed to be \cite{Tian:2016uwp,PhysRevLett.113.020403,Menezes:2017akp,Dicke:1954zz},
\begin{equation}
H_{int}(\tau)=\lambda\left[\sigma^{(1)}_{2} \Phi({\bf 
x}_1(\tau))+\sigma_{2}^{(2)}\Phi({\bf x}_2(\tau))\right]~,
\end{equation}
where $\lambda$ is the coupling constant which is taken to be small. The final results are presented in the paper will remain identical with $\sigma_1$ instead of $\sigma_2$ in the interaction Hamiltonian.

Initially, we assume 
that there was no interaction between the atoms and the external field 
\cite{Tian:2016uwp} and also there is no correlated state
between the atoms and quantum field through other medium so the total density matrix of the system can be expressed as  
$\rho_{t}(0)=\rho(0)\otimes|0\rangle\langle0|$. Here $\rho(0)$ is the initial density matrix of the 
two-atom system and $|0\rangle$ is the vacuum state of the scalar field. In the 
frame of atoms which is also considered as a proper frame, the time evolution of 
the density matrix of the total system obeys von Neumann equation 
\cite{Tian:2016uwp} \emph{i.e.}
\begin{equation}\label{rho_g}
 \frac{\partial \rho_{t}(\tau)}{\partial 
\tau}=-i\left[H(\tau),\rho_{t}(\tau)\right]~,
\end{equation}
where $\tau$ is the proper time. Here we assumed that the two atoms should be closely spaced so that they should
not experience any relative curvature effect. Otherwise, the proper time carried by the clocks
of the atoms will differ by a gravitational redshift factor then equation (\ref{rho_g}) is not viable. We are interested in the time evolution of the 
two-atom system and in order  to obtain the reduced dynamics, we trace out the 
field degrees of freedom \emph{i.e.} $\rho(\tau)= Tr_{F}[\rho_{t}(\tau)]$. In 
the weak-coupling limit, the resulting equation then becomes the 
Kossakowski-Lindblad form \cite{Lindblad1976,Tian:2016uwp,1976JMP....17..821G} 
which is given by,
\begin{equation} \label{rho}
 \frac{\partial \rho (\tau)}{\partial 
\tau}=-i[H_{e},\rho(\tau)]+L[\rho(\tau)]~,
\end{equation}
where $H_{e}$ is the effective Hamilton of the two-atom system which is given by 
\cite{Tian:2016uwp},
\begin{equation}\label{effectivehamilton}
H_{e}= H_{A} -\frac { i } {2}\sum^{2}_{a,b=1}\sum^{3}_{j,k=1}H^{ (ab) } _ { 
jk }\sigma^{(a)}_{j}\sigma^{(b)}_{k}~.
\end{equation}
Here we do not bother about the form of $L[\rho(\tau)]$ in equation (\ref{rho}) 
because it is not important in our calculation \cite{Tian:2016uwp}. The 
elements, $H^{ (ab) } _ { jk },$ in the the effective Hamilton, $H_{e}$ 
(\ref{effectivehamilton}), can be determined by the Hilbert transforms of the 
two-point functions, $\mathcal{K}^{(ab)}(\Omega)$, which are given by 
\cite{Tian:2016uwp},
\begin{equation}\label{hilbert}
 \mathcal{K}^{(ab)}(\Omega)=\frac{P}{\pi i}\int^{\infty}_{-\infty} d 
\omega \frac{\mathcal{{G}}^{(ab)}(\omega)}{\omega-\Omega}~,
\end{equation}
where P is the principal value of the integral. Here $a,b$ are dummy variables which can take the 
value 1 or 2 and $\mathcal{{G}}^{(ab)}(\Omega)$ denote the Fourier transforms of 
the two-point functions which are given by, 
\begin{equation}\label{twoFourier}
 \mathcal{{G}}^{(ab)}(\Omega)=\int^{\infty}_{-\infty} d \Delta \tau 
~e^{i\Omega \Delta \tau}~{G}^{(ab)}(\Delta\tau)~,
\end{equation}
where ${G}^{(a b)}(\Delta \tau)$ indicate the two-point functions which are 
given by,
\begin{equation}\label{twopoint}
{G}^{(a b)}(\Delta \tau)=\langle{\Phi}(\tau,\xx_{a}){\Phi}(\tau',\xx_{b})\rangle~,
\end{equation}
and $\Delta \tau = (\tau-\tau')$. Then the explicit forms of the elements, $H^{ 
(ab) } _ { jk }$, can be written as \cite{Tian:2016uwp}, 
\begin{equation}\label{H}
 H^{ (ab) } _ { jk }=A^{(a b)}\delta_{jk}-i B^{(a 
b)}\epsilon_{jkl}\delta_{3 l}-A^{(a b)}\delta_{3j}\delta_{3k}~,
\end{equation}
where the parameters $A^{(a b)}$ and $B^{(a b)}$ are given by,
\begin{eqnarray}\label{AB}
 A^{(a b)}&=&\frac{\lambda^2}{4}\left[\mathcal{K}^{(ab)} 
(\omega_0)+\mathcal{K}^{(ab)}(-\omega_0)\right]\label{a}~,\nonumber\\ 
B^{(ab)}&=&\frac{\lambda^2}{4}\left[\mathcal{K}^{(ab)} 
(\omega_0)-\mathcal{K}^{(ab)}(-\omega_0)\right]\label{b}~.
\end{eqnarray}
It has been already mentioned that the first term \emph{i.e.} $H_{A}$ of the 
effective Hamiltonian, $H_{e}$ (\ref{effectivehamilton}), is the total 
Hamilton of two isolated atoms. The last term of the effective Hamiltonian 
(\ref{effectivehamilton}) \emph{i.e.} 
\begin{equation}
H_{LS}\equiv-\frac{i}{2}\sum^{2}_{a,b=1}\sum^{3}_{j,k=1}H^{ (ab) } _ { jk 
}\sigma^{(a)}_{j}\sigma^{(b)}_{k},
\end{equation}
plays the identical role as the Lamb shift of the two-atom system which 
arises due to the interaction between the atoms and the external field 
\cite{Tian:2016uwp}.

\subsection{The shifts of the energy level of the two-atom system} It is shown 
that when both the atoms are in the ground state or the excited state, there 
exists no interatomic correlation between them. Interatomic correlations only 
exist in the symmetric and antisymmetric states (entangled states) 
\citep{Tian:2016uwp}. In this work, we want to investigate the effect of 
interatomic correlations between two atoms. We achieve this by calculating the 
shifts in the energy levels of the symmetric and antisymmetric states of the 
entangled two-atom system. By calculating the average values of $H_{LS}$ on the 
symmetric state 
$|S\rangle=\frac{1}{\sqrt{2}}(|e_{1}\rangle|g_{2}\rangle+|g_{1}\rangle|e_{2} 
\rangle)$ and the antisymmetric state 
$|A\rangle=\frac{1}{\sqrt{2}}(|e_{1}\rangle|g_{2}\rangle-|g_{1}\rangle|e_{2} 
\rangle)$, one can obtain the shifts of the energy level of the symmetric state 
and the antisymmetric state  as \cite{Tian:2016uwp},
\begin{eqnarray}\label{energyshifts}
\delta E_{S_{LS}}&=&\langle
S|H_{LS}|S\rangle=-\frac{i}{2}\left[\sum^{3}_{j=1}\left(H^{12}_{
jj}+H^{21}_{jj}+H^{11}_{jj}+H^{22}_{jj}\right)-2(H^{12}_{33}+H^{ 21 } _ { 33 } 
) \right],
\nonumber\\
\delta E_{A_{LS}}&=&\langle
A|H_{LS}|A\rangle=\frac{i}{2}\left[\sum^{3}_{j=1}\left(H^{12}_{
jj}+H^{21}_{jj}-H^{11}_{jj}-H^{22}_{jj}\right)\right].
\end{eqnarray}
Here $\delta E_{S_{LS}}$ stands for the first-order energy level shift of the 
symmetric state and $\delta E_{A_{LS}}$ stands for corresponding energy level 
shift of the antisymmetric state. We note that these shifts of the energy level 
are related to the two-point functions through the equations 
(\ref{hilbert})-(\ref{b}) which can be computed along the trajectories of the 
atoms and thus they depend on the spacetime background. As a result, these 
shifts of the energy level are different in different spacetimes. Here we 
compute these shifts of the energy level for the Schwarzschild spacetime when 
the atoms are located close to the horizon and these shifts when the atoms are 
in a thermal Minkowski spacetime.

\section{\emph{RCPI} for the Schwarzschild spacetime when two static entangled 
atoms are located close to the horizon}\label{Schwarzschild spacetime}

\subsection{Schwarzschild metric in near horizon region}

Here we consider (3+1) dimensional Schwarzschild spacetime to compute the 
\emph{RCPI} between two entangled atoms when the atoms are located close to the 
horizon. The Schwarzschild spacetime is described by the metric
\begin{equation}\label{SchwarzschildMetric01}
d{s}^2 = - f(r) dt^2 + f(r)^{-1} dr^2 
+ r^2 d\theta^2 + r^2 \sin^2\theta d\phi^2 ~,
\end{equation}
where $f(r) = \left(1- r_s /r\right)$ and $r_s = 2 G M$ is the Schwarzschild 
radius related to the metric. A proper distance from the horizon 
to a radial distance $r$ is defined by the formula \cite{Ydri:2017oja},
\begin{eqnarray}
&&l= \int^r_{r_s}\frac{d r'}{\sqrt{1-\frac{r_s}{r'}}}\nonumber\\
&&=\sqrt{r(r-r_s)}+r_s \sinh^{-1}(\sqrt{\frac{r}{r_s}-1})~.
\end{eqnarray}
In terms of $l$, the Schwarzschild metric (\ref{SchwarzschildMetric01}) becomes
\begin{equation}\label{SchwarzschildMetric02}
d{s}^2 = - f(r) dt^2 + d l^2
+ r^2(l) d\theta^2 + r^2(l) \sin^2\theta d\phi^2 ~,
\end{equation}
where $f(r)= \left(1- r_s /r(l)\right)$. Now near the horizon, where 
$r=r_s+\delta$  and $l=2 \sqrt{r_s\delta}$,  within the leading order 
approximation, the Schwarzschild metric (\ref{SchwarzschildMetric02}) becomes 
\cite{Polchinski:2016hrw,Ydri:2017oja,Dabholkar:2012zz, 
Goto:2018ijz}
\begin{equation}\label{rindler}
 d{s}^2 = -l^2 \frac{dt^2}{4 r_s^2}+d l^2+r_s^2 
d\theta^2+r_s^2\sin^2\theta d\phi^2~.
\end{equation}
Here we have considered $\delta$ is a positive parameter and $\delta << \rs$. 
Now we can define new coordinates which are given by,
\begin{equation}\label{Kruskal}
X_1=l\cosh\frac{t}{2 r_s}~;~~ T=l\sinh\frac{t}{2 r_s}~.
\end{equation}
Using these coordinates (\ref{Kruskal}), the metric (\ref{rindler}) becomes
\begin{equation}\label{Minkowski}
 d{s}^2=-d T^2+d X_1^2+r_s^2 
d\theta^2+r_s^2\sin^2\theta d\phi^2~.
\end{equation}
If we only focus on a small angular region near the horizon which is around 
$\theta= 0$ then we can replace the angular coordinates with the Cartesian 
coordinates which are given by \cite{Ydri:2017oja, Goto:2018ijz}
\begin{equation}\label{Cartesian}
 X_2= r_s\theta \cos\phi~;~~X_3=r_s \theta \sin\phi~.
\end{equation}
Using these coordinates (\ref{Cartesian}), the equation (\ref{Minkowski}) 
becomes
\begin{equation}\label{Minkowski1}
 d{s}^2=-d T^2+d X_1^2+d X_2^2+d X_3^2~,
\end{equation}
which expresses the Minkowski spacetime \cite{Ydri:2017oja}.

\subsection{\emph{Resonance Casimir-Polder interaction}}

In the position space, using the coordinates of the inertial metric 
(\ref{Minkowski1}), the two-point function for a massless scalar field can be 
expressed as
\begin{equation}\label{Two-point}
G(x,x') \equiv \langle 0|\hat{\Phi}(x) \hat{\Phi}(x')|0\rangle
= \langle 0|\hat{\Phi}(T,\x) \hat{\Phi}(T',\x')|0\rangle~,
\end{equation}
where $|0\rangle$ indicates corresponding vacuum state. Here we ignore 
back-reaction of this scalar field on the spacetime metric. It can be shown that 
in Fock quantization, the two-point function (\ref{Two-point}) becomes in this 
usual form \cite{Birrell1984quantum,Sriramkumar:1994pb},
\begin{equation}\label{Twopoint_inertial}
 G(x,x')=-\frac{1}{4 \pi^2} \frac{1}{(\Delta T-i 
\epsilon)^2 -\lvert\Delta \x\rvert^2}~,
\end{equation}
where  $-\Delta T^2+\lvert\Delta \x\rvert^2= 
-(T-T')^2+(X_1-X_1')^2+(X_2-X_2')^2+(X_3-X_3')^2$ is  the Lorentz invariant spacetime 
interval and $\epsilon$ is a small, positive parameter which is introduced to 
evaluate two-point function.

Now we assume that two static atoms are located at the position 
($r,\theta,\phi$) and ($r,\theta',\phi$) which are close to the horizon and the 
angles $\theta$ and $\theta'$ are taken to be small. Using the equations 
(\ref{twopoint},\ref{Kruskal},\ref{Cartesian},\ref{Twopoint_inertial}), we 
obtain the two-point functions for these two spacetime points which are 
given by,
\begin{eqnarray}\label{Gaa}
\nonumber
&&G^{(11)}(x, 
x^\prime)=G^{(22)}(x,x')\\ \nonumber
&&=-\frac{1}{4\pi^2}\bigg[(l\sinh\,
t/2 \rs-l\sinh\,t^\prime/2 \rs -i\epsilon)^2
\\  \nonumber
&&-(l\cosh\,t/2 \rs-l\cosh\,
t^\prime/2 \rs)^2\bigg]^{-1}
\\ 
&=&-\frac{1}{16\pi^2l^2\sinh^2(\frac{\Delta\tau}{2l}-i\epsilon)}~,
\end{eqnarray}
and
\begin{eqnarray}\label{Gab}
\nonumber
&&G^{(12)}(x, x')=G^{(21)}(x, x')
\\   \nonumber
&=&-\frac{1}{4\pi^2}\bigg[(l\sinh\,t/2 \rs-l\sinh\,t^\prime/2 \rs-i\epsilon)^2
\\  \nonumber
&&-(l\cosh\,t/2 \rs-l\cosh\,
t^\prime/2 \rs)^2
\\  \nonumber
&&-(\rs \theta\cos\phi-\rs \theta^\prime\cos\phi)^2
\\  \nonumber
&&-(\rs\theta\sin\phi-\rs
\theta^\prime\sin\phi)^2\bigg]^{-1}
\\ 
&=&-\frac{1}{16\pi^2l^2}\frac{1}{\sinh^2(\frac{\Delta\tau}{2l}
-i\epsilon)-\frac{\rs^2}{l^2}(\frac{\Delta\theta}{2})^2},
\end{eqnarray}
where $\Delta\tau=\frac{l}{2\rs}(t-t')=\frac{l}{2\rs}\Delta t$ with $\tau$ which 
is the proper time of the static atoms in the Schwarzschild spacetime and 
$\Delta\theta=(\theta-\theta')$. Here we have considered the atoms are located at $(r, \theta, \phi)$
and $(r,\theta' ,\phi)$ respectively, where r is close to the horizon. One can consider the other cases where the two atoms located at
the same $(\theta, \phi)$ value but on the two sides of the horizon and
at $(r, \theta, \phi)$ and $(r, \theta, \phi' )$. The final results will be different for both the cases because the two-point functions will be different for those cases.  Now using this two-point functions 
(\ref{Gaa},\ref{Gab}), we can compute the Fourier transforms 
(\ref{twoFourier}) of these two-point functions which are given by,
\begin{eqnarray}\label{Fourier1}
&&\mathcal{{G}}^{(11)}(\Omega)=\mathcal{{G}}^{(22)}
(\Omega)\nonumber\\&&=\int^{
\infty } _ { -\infty
} -\frac{1}{16 
\pi^2l^2~\sinh^{2}(\frac{\Delta\tau}{2l}-i\epsilon)}~e^{i\Omega \Delta \tau}
d \Delta \tau\nonumber\\
&&=\frac{1}{2 \pi}\frac{\Omega}{1-e^{-2 \pi l\Omega}}~,
\end{eqnarray}
and
\begin{eqnarray}\label{Fourier2}
&&\mathcal{{G}}^{(12)}(\Omega)=\mathcal{{G}}^{(21)}
(\Omega)\nonumber\\&&=\int^{
\infty } _ { -\infty} 
-\frac{1}{16\pi^2l^2}\frac{1}{\sinh^2(\frac{\Delta\tau}{2l}
-i\epsilon)-\frac{\rs^2}{l^2}(\frac{\Delta\theta}{2})^2}~e^{i\Omega \Delta \tau}
d \Delta \tau\nonumber\\
&&=\frac{1}{2 \pi}\frac{\Omega}{1-e^{-2 \pi l\Omega}}~g(\Omega,L/2),
\end{eqnarray}
where we define $g(\Omega,z)=\frac{\sin\left[2 l \Omega \sinh^{-1}(z/ 
l)\right]}{2 z \Omega \sqrt{1+z^2/ l^2}}$. Within the leading order 
approximation, here $L=\rs(\Delta \theta)$ denotes the proper distance between 
the two points ($r,\theta,\phi$) and ($r,\theta',\phi$) which are close to the 
horizon where the angles $\theta$ and $\theta'$ are taken to be small. Now using 
the Fourier transforms of two-points functions (\ref{Fourier1},\ref{Fourier2}), 
we can compute the Hilbert transforms of the two-point functions 
(\ref{hilbert}) which are given by,
\begin{eqnarray}
\mathcal{K}^{(11)}(\omega_{0})&=&\mathcal{K}^{(22)}(\omega_{0}
)\nonumber\\&=&\frac { P } { 2 
\pi^2 i}\int^{\infty}_{-\infty} d 
\omega \frac{1}{\omega-\omega_0}\frac{\omega}{1-e^{-2 \pi 
l\omega}}~,\nonumber
\end{eqnarray}
and
\begin{eqnarray}
\mathcal{K}^{(12)}(\omega_{0})&=&\mathcal{K}^{(21)}(\omega_{0}
)\nonumber\\&=&\frac { P } { 2 
\pi^2 i}\int^{\infty}_{-\infty} d 
\omega \frac{1}{\omega-\omega_0}\frac{\omega}{1-e^{-2 \pi 
l\omega}}~g(\omega,L/2)\nonumber~.
\end{eqnarray}
After putting these Hilbert transforms into the equations (\ref{H}) and 
(\ref{AB}), we get
\begin{eqnarray}
 H^{ (11) } _ { jk }=H^{ (22) } _ { jk } =A_{1}\delta_{jk}-i 
B_{1}\epsilon_{jkl}\delta_{3 l}-A_{1}\delta_{3j}\delta_{3k}~,\nonumber\\
H^{ (12) } _ { jk }=H^{ (21) } _ { jk } =A_{2}\delta_{jk}-i 
B_{2}\epsilon_{jkl}\delta_{3 l}-A_{2}\delta_{3j}\delta_{3k}~,\label{HR}
\end{eqnarray}
where the parameters $A_{1}$, $B_{1}$, $A_{2}$ and $B_{2}$ are given by,
\begin{eqnarray}
A_{1}=\frac{\lambda^2 P}{8 \pi^2 i}\int^{\infty}_{-\infty} d\omega 
\left(\frac{\omega}{\omega-\omega_0}+\frac{\omega}{\omega+\omega_0}\right)\frac{
1 } { 1-e^{-2 \pi l\omega}}~,\nonumber\\
B_{1}=\frac{\lambda^2 P}{8
\pi^2 i}\int^{\infty}_{-\infty} d 
\omega 
\left(\frac{\omega}{\omega-\omega_0}-\frac{\omega}{\omega+\omega_0}\right)\frac{
1 } {
1-e^{-2 \pi 
l\omega}}~,\nonumber\\
A_{2}=\frac{\lambda^2 P}{8
\pi^2 i}\int^{\infty}_{-\infty} d 
\omega 
\left(\frac{\omega}{\omega-\omega_0}+\frac{\omega}{\omega+\omega_0}\right)\frac{
1 } {
1-e^{-2 \pi 
l\omega}}\nonumber\\\times~g(\omega,L/2)~,\nonumber\\
B_{2}=\frac{\lambda^2 P}{8
\pi^2 i}\int^{\infty}_{-\infty} d 
\omega 
\left(\frac{\omega}{\omega-\omega_0}-\frac{\omega}{\omega+\omega_0}\right)\frac{
1 } {
1-e^{-2 \pi 
l\omega}}\nonumber\\\times~g(\omega,L/2)~.\label{AB1}
\end{eqnarray}
Using the equations (\ref{HR}) and (\ref{AB1}), one can calculate the shifts of 
the energy level of the symmetric state and the antisymmetric state of the 
two-atom system (\ref{energyshifts}) which are given by,
\begin{eqnarray}\label{energy}
\delta E_{S_{LS}}= -\frac{\lambda^2 }{4
\pi^2 }\int^{\infty}_{0} d 
\omega\left(\frac{\omega}{\omega-\omega_0}+\frac{\omega}{\omega+\omega_0}
\right)\left[ g(\omega,L/2)+1\right]~,\nonumber\\
\delta E_{A_{LS}}=\frac{\lambda^2 }{4
\pi^2 }\int^{\infty}_{0} d 
\omega\left(\frac{\omega}{\omega-\omega_0}+\frac{\omega}{\omega+\omega_0}
\right)
\left[ g(\omega,L/2)-1\right]~.
\end{eqnarray}
From the above equations (\ref{energy}), it is shown that the shifts of the 
energy level of the symmetric state and the antisymmetric state depend on the 
proper distance, $L$, between the atoms. So the interatomic interactions exist 
in the symmetric state and the antisymmetric state of the two-atom system 
\cite{Tian:2016uwp}. Again for computing \emph{Casimir-Polder force} between the 
two atoms, one has to take the derivative with respect to $L$. So one can 
neglect the terms which do not depend on $L$ from the above equations 
(\ref{energy}) to rewrite the interatomic interactions \cite{Tian:2016uwp}.  The 
interatomic interactions for the symmetric state and the antisymmetric state of 
the two-atom system are then given by,
\begin{eqnarray}\label{energy1}
\delta E_{S}&=& -\frac{\lambda^2 }{4
\pi^2 }\int^{\infty}_{0} d 
\omega\left(\frac{\omega}{\omega-\omega_0}+\frac{\omega}{\omega+\omega_0}\right)
 g(\omega,L/2)~,\nonumber\\
\delta E_{A}&=&\frac{\lambda^2 }{4
\pi^2 }\int^{\infty}_{0} d 
\omega\left(\frac{\omega}{\omega-\omega_0}+\frac{\omega}{\omega+\omega_0}\right)
g(\omega,L/2).
\end{eqnarray}
We can evaluate the integral in the above equations (\ref{energy1}) analytically
and the results are given by,
\begin{eqnarray}\label{interatomicinteraction}
\delta E_{S}&=& -\frac{\lambda^2 }{4
\pi L\sqrt{1+(L/2l)^2} }\cos(2 \omega_0l
\sinh^{-1}(L/2l))~,\nonumber\\
\delta E_{A}&=&\frac{\lambda^2 }{4
\pi L\sqrt{1+(L/2l)^2} }\cos(2 \omega_0l
\sinh^{-1}(L/2l))~.
\end{eqnarray}
These are the \emph{resonance interatomic interactions} between two entangled 
atoms for the Schwarzschild spacetime when the atoms are located close to the 
horizon. 

We note that the characteristic length scale $l$ which up to the leading order 
approximation is equal to  $\frac{ \sqrt{1-\frac{r_s}{r}}}{\kappa}$, where 
$\kappa=\frac{1}{2 \rs}$ is the surface gravity. In order to investigate the 
detailed behavior of the \emph{RCPI} between two entangled atoms widely, here we 
consider both the limits of the proper distance between the atoms which are 
larger and smaller than the characteristic length scale. When the proper 
distance between two atoms is larger than the characteristic length scale, the 
metric shows a strong noninertial character where the results should be 
different with that corresponding to the Minkowski spacetime. Whereas, when the 
proper distance between two atoms is smaller than the characteristic length 
scale, it is possible to find a local inertial frame where the results should be 
the same, as obtained in Minkowski spacetime.

From equation (\ref{interatomicinteraction}), it is shown that in the limit 
$L>>l$ (or $\Delta\theta>> 2 \sqrt{\frac{\delta}{r_s}}$), the \emph{RCPI} can be 
expressed as 
\begin{eqnarray}
\label{interatomicinteraction1}
\delta E_{S}= -\frac{\lambda^2~l}{2
\pi L^2 }\cos(2 \omega_0l
\log(L/l))~,\nonumber\\
\delta E_{A}=\frac{\lambda^2~l }{2
\pi L^2 }\cos(2 \omega_0l
\log(L/l))~,
\end{eqnarray}
and in the limit $L<<l$ (or $\Delta\theta<< 2 \sqrt{\frac{\delta}{r_s}}$), the 
\emph{RCPI} can be expressed as
\begin{eqnarray}
\label{interatomicinteraction2}
\delta E_{S}= -\frac{\lambda^2 }{4
\pi L }\cos( \omega_0 L)~,\nonumber\\
\delta E_{A}=\frac{\lambda^2  }{4
\pi L }\cos( \omega_0 L)~.
\end{eqnarray}
In this limit, the effective spacetime curvature is neglected so the result is same as obtained in thermal Minkowski spacetime.

Here we have shown that beyond a characteristic length scale which is 
proportional to the inverse of the surface gravity $\kappa$, the \emph{RCPI} 
between two entangled atoms is characterized by a $1/L^2$  power-law decay for 
the Schwarzschild spacetime when the atoms are located close to the horizon.  We 
have also shown that beyond the characteristic length scale, the \emph{RCPI} 
depends on the characteristic length scale which is also related to 
redshifted Unruh-Davies temperature temperature 
$T=1/2 \pi l=\frac{\kappa}{2 \pi k_B \sqrt{1-\frac{r_s}{r}}}$ measured by a 
static observer near the horizon \cite{Singleton:2011vh,Smerlak:2013sga,ISRAEL1976107,Louko:1998dj, ISRAEL1981204,Hinton:1982, 
Birrell1984quantum, 
Hodgkinson:2014iua,Acquaviva:2011vq,Acquaviva:2013oaa,Sriramkumar:1994pb, 
book:250016}. This temperature is equal to the Hawking 
temperature \cite{hawking1975,Barman:2017fzh,Lambert:2013uaa, Jacobson:2003vx, Smerlak:2013sga,ISRAEL1976107,ISRAEL1981204,
Kiefer:2002fp, Traschen:1999zr, DEWITT1975295, Ford:1997hb, 
Hollands:2014eia,Padmanabhan:2009vy, 
Chakraborty:2015nwa,Chakraborty:2017pmn,Helfer:2003va,Carlip:2014pma, 
Fulling1987135,Hinton:1982, Parikh:1999mf,Visser:2001kq, Singleton:2011vh, 
Birrell1984quantum, Louko:1998dj,
Hodgkinson:2014iua,Acquaviva:2011vq,Acquaviva:2013oaa,Sriramkumar:1994pb, 
book:250016,Bhattacharya:2013tq,Lapedes:1977ip, Davies:1974th, 
Wald1975,Singh:2014paa,Jacobson:2012ei, Hartle:1976tp} 
 measured by the observer.

\section{Comparing the relative proper acceleration between two atoms and \emph {RCPI} in Schwarzschild spacetime} 
As Schwarzschild spacetime is curved spacetime so for an atom located with a radial distance $r$ from the black hole,
its proper acceleration is $a=\frac{M}{ r^2\sqrt{1-\rs/r}}$, then the relative proper 
acceleration of the two atoms located with the same radial distance $r$ from the black hole
with an azimuthal interval $\Delta 
\theta=\theta-\theta'<<1$ follows 
\begin{equation}
\Delta a= 2 a \sin (\Delta \theta/2)\simeq a \Delta \theta= \frac{M}{r^2\sqrt{1-\rs/r}}\Delta\theta.
\end{equation}
Now relative acceleration for the atoms located near the horizon becomes,
\begin{equation} \Delta 
a|_{r\rightarrow\rs}\simeq\frac{M}{ r\sqrt{r(r-\rs)}}\Delta 
\theta\simeq\frac{\Delta \theta}{2\sqrt{\rs \delta}}=\frac{\Delta \theta}{l}. 
\end{equation} 
So the interatomic separation $L>>l$ (or $\Delta\theta>> 2 
\sqrt{\frac{\delta}{r_s}}$) is equivalent to $\Delta a|_{r\rightarrow\rs}>> 
\frac{1}{\rs}$, indicating that the relative proper acceleration of the two 
atoms in the case of $L >> l$ is not negligible, and thus the dynamics of the 
two atoms can not be depicted in the same local inertial frame. So we can 
take $L>>l$ limit as a result Eq. (\ref{interatomicinteraction1}) is not valid.  But the relative proper acceleration of the two 
atoms in the case of $L << l$ is  negligible, and thus the dynamics of the 
two atoms can be depicted in the same local inertial frame. So we can
take $L<<l$ limit as a result Eq. (\ref{interatomicinteraction2}) is valid.
 We have shown that, for Schwarzschild spacetime, beyond a characteristic length scale which is 
proportional to the inverse of the surface gravity $\kappa$, the \emph{Resonance 
Casimir-Polder interaction (RCPI)} between two entangled atoms is characterized 
by a $1/L^2$  power-law provided the atoms are located close to the horizon. 
But here we have shown that the length scale limit beyond a characteristic value is not valid for Schwarzschild spacetime.

\section{\emph{RCPI} for the thermal Minkowski spacetime}\label{Minkowski 
spacetime}

In order to compare the \emph{RCPI} behavior for thermal Minkowski spacetime and Schwarzschild spacetime  Minkowski spacetime here  we have considered two static atoms in Minkowski spacetime. 
These atoms are coupled to a massless scalar field in a thermal state with the 
temperature $T=1/2 \pi l$.  The two-point functions for this case  are given by 
\cite{Birrell1984quantum,Tian:2016uwp},
\begin{eqnarray}
 {G}^{(11)}(x,x')&=&{G}^{(22)}(x,x')\nonumber\\&=&- \frac{1}{4 
\pi^2}\sum^{+\infty}_{m=-\infty}\frac{1}{(\Delta \tau- 
im/T-i\epsilon)^2}\nonumber
\end{eqnarray}
and
\begin{eqnarray}
{G}^{(12)}(x,x')&=&{G}^{(21)}(x,x')\nonumber\\&=&- \frac{1}{4 
\pi^2}\sum^{+\infty}_{m=-\infty}\frac{1}{(\Delta \tau- 
im/T-i\epsilon)^2-L^2}~,\nonumber
\end{eqnarray}
where $ \Delta \tau =​t-​t'$ with t which is the proper time of the static atoms 
in Minkowski spacetime and $L=2r\sin(\Delta\theta/2)$ is the Euclidean distance 
between them when the atoms are located at the position ($r,\theta,\phi$) and 
($r,\theta',\phi$) in this spacetime \cite{Tian:2016uwp}. From these two-point 
functions, one can compute the \emph{RCPI} between two entangled atoms for the 
thermal Minkowski spacetime using the similar method. The interatomic 
interactions for the symmetric state and the antisymmetric state for this 
spacetime look like \cite{Tian:2016uwp},
\begin{gather}
\delta E_{S_{M}}= -\frac{\mu^2 }{4
\pi L }\cos( \omega_0 L)~,\nonumber\\
\delta E_{A_{M}}=\frac{\mu^2  }{4
\pi L }\cos( \omega_0 L)~.
\end{gather}
Here we have shown that the \emph{RCPI} between two entangled atoms is always 
characterized by a $1/L$ power-law decay for the Minkowski spacetime \cite{Tian:2016uwp}.  We 
have also shown that the \emph{RCPI} for this spacetime is always 
temperature-independent \cite{Tian:2016uwp} and similar to the \emph{RCPI} 
between two inertial atoms shown in the equation 
(\ref{interatomicinteraction2}). In Minkowski spacetime the surface gravity $\kappa$ is  zero so always $L<<l$ and thus the
final result is  always independent of the temperature and hence of the associated scale $l$, no
matter whether the field state is thermal or not \cite{Tian:2016uwp}.
\bigskip
\section{Discussion}

Here we have applied the open quantum system approach to obtain effective 
Hamiltonian of two atoms. This effective Hamiltonian allow us to compute the 
\emph{RCPI} between two entangled atoms. Subsequently, we have calculated the 
\emph{RCPI} for the Schwarzschild spacetime when the atoms are located close to 
the horizon. Although we have shown that beyond a characteristic length scale 
which is proportional to the inverse of the surface gravity $\kappa$, the 
\emph{RCPI} is characterized by a $1/L^2$  power-law decay for the Schwarzschild 
spacetime. We have also shown that beyond the characteristic length scale, the 
\emph{RCPI} for this spacetime depends on the characteristic length scale which 
is also related to the temperature measured by a static observer near the 
horizon.  Whereas, the \emph{RCPI} is temperature-independent and is  always 
characterized by a $1/L$ power-law decay for the thermal Minkowski spacetime. Seemingly, it appears that the spacetimes can be distinguished from each other using the \emph{RCPI} behavior.
But our further exploration leads to the conclusion that the length scale limit beyond a characteristic value is not valid for Schwarzschild spacetime.    
In summary, using the \emph{RCPI} between two entangled atoms, one can not 
distinguish these two spacetimes.

\emph{acknowledgments.}-

I thank Golam Mortuza Hossain, Sumanta Chakraborty, Arnab Chakrabarti and Gopal Sardar for many 
useful discussions. I also thank IISER Kolkata for supporting this work through 
doctoral fellowships.

\end{document}